\documentclass[twocolumn,showpacs,showkeys,prc,aps,preprintnumbers,superscriptad
dress]{revtex4}

\usepackage{graphicx}
\usepackage{dcolumn}
\usepackage{bm}

\begin{document}
\title{ Beam-Energy and System-Size Dependence of Dynamical Net Charge 
Fluctuations}

\author{B.~I.~Abelev}\affiliation{University of Illinois at Chicago, Chicago, 
Illinois 60607, USA}
\author{M.~M.~Aggarwal}\affiliation{Panjab University, Chandigarh 160014, India}
\author{Z.~Ahammed}\affiliation{Variable Energy Cyclotron Centre, Kolkata 
700064, India}
\author{B.~D.~Anderson}\affiliation{Kent State University, Kent, Ohio 44242, 
USA}
\author{D.~Arkhipkin}\affiliation{Particle Physics Laboratory (JINR), Dubna, 
Russia}
\author{G.~S.~Averichev}\affiliation{Laboratory for High Energy (JINR), Dubna, 
Russia}
\author{Y.~Bai}\affiliation{NIKHEF and Utrecht University, Amsterdam, The 
Netherlands}
\author{J.~Balewski}\affiliation{Massachusetts Institute of Technology, 
Cambridge, MA 02139-4307, USA}
\author{O.~Barannikova}\affiliation{University of Illinois at Chicago, Chicago, 
Illinois 60607, USA}
\author{L.~S.~Barnby}\affiliation{University of Birmingham, Birmingham, United 
Kingdom}
% ...

\collaboration{STAR Collaboration}\noaffiliation

\begin{abstract}
We present measurements of net charge  fluctuations in $Au + Au$ collisions at 
$\sqrt{s_{NN}} = $ 19.6, 62.4, 130, 
and 200 GeV, $Cu + Cu$ collisions at $\sqrt{s_{NN}} = $ 62.4, 200 GeV, and $p + 
p$ 
collisions at $\sqrt{s} = $ 200 GeV using the dynamical net charge fluctuations 
measure  $\nu_{+-{\rm,dyn}}$. 
We observe that the dynamical fluctuations are non-zero at all energies and 
exhibit a modest dependence on beam 
energy. A weak system size dependence is also observed. We examine the collision 
centrality dependence of the net 
charge fluctuations and find that dynamical net charge fluctuations violate 
$1/N_{ch}$ scaling, but display 
approximate $1/N_{part}$ scaling. We also study the azimuthal and rapidity 
dependence of 
the net charge correlation strength and observe strong dependence on the 
azimuthal angular range and pseudorapidity
 widths integrated to measure the correlation. 
\end{abstract}


\maketitle
\setcounter{page}{1}

\section{Introduction}\label{intro}

Anomalous transverse momentum and net charge event-by-event fluctuations have 
been proposed as indicators of the 
formation of a quark gluon plasma (QGP) in the midst of high-energy heavy ion 
collisions. A number of authors 
\cite{JeonKoch00,Heiselberg01,Asakawa00} have argued that entropy conserving 
hadronization of a plasma of quarks and
 gluons should produce a final state characterized by a dramatic reduction of 
the net charge fluctuations relative 
to that of a hadron gas. Simply put, their prediction relies on the notion that 
quark-quark correlations can be
 neglected, and hadronization of gluons produces pairs of positive and negative 
particles not contributing to the 
net charge fluctuations. Accounting for the fractional charge of the quarks, 
they find that, for a QGP, the variance of 
the ratio of positive and negative particles scaled by the total charged 
particle multiplicity, a quantity they 
call $D$, should be approximately four times smaller than for a gas of hadron. 
Quark-quark correlation may 
however not be negligible; Koch {\em et al.} \cite{JeonKoch00} extended their 
original estimates to include 
susceptibilities calculated on the lattice. They find the quantity 
$D=4\langle\!\Delta Q^2\!\rangle/N_{ch}$ (where, $\Delta Q^2$ 
is the variance of the net charge, Q = $N^{+}$ - $N^{-}$ and $N_{ch}$ is the 
total number of charged particles 
observed in a particular momentum space window under consideration) is 
quantitatively 
different from their first basic estimate but nonetheless still dramatically 
smaller than values expected for a 
hadron gas. It is clear however that hadron collisions, and in particular heavy 
ion collisions, produce substantial 
number of many high mass particles, 
and specifically short lived (neutral) particles or resonances which decay into 
pairs of positive and negative 
particles. Such decays increase the multiplicity of charged particles in the 
final state while
 producing negligible impact on the net charge variance. Jeon and Koch have in 
fact argued that one can use the 
magnitude of net charge fluctuations to estimate the relative production of 
$\rho$ and $\omega$ mesons 
\cite{JeonKoch99}. Calculations based on a thermal model lead to a value $D$ of 
order of 2.8  which although reduced 
relative to the value expected for a pion gas is nonetheless remarkably larger 
than that predicted for a QGP 
\cite{JeonKoch00}. Note that transport models such as UrQMD predict values in 
qualitative 
agreement with those of thermal models \cite{Bleicher00}. A measurement of net 
charge fluctuations therefore appears,
 on the outset, as an interesting means to 
identify the formation of quark gluon plasma in high-energy heavy ion 
collisions.

First measurements of net charge fluctuations were reported by both PHENIX 
\cite{AdcoxPRC89} and 
STAR \cite{Adams03c} collaborations on the basis of $Au + Au$ data acquired 
during the first RHIC run at 
$\sqrt{s_{NN}} =$ 130 GeV. Measurements were reported by PHENIX 
\cite{AdcoxPRC89} in terms of a 
reduced variance, $\omega_Q$ = $\langle\!\Delta Q^{2}\!\rangle$/$N_{ch}$. 
Unfortunately, measured values of this quantity 
depend on the efficiency.
STAR instead reported results \cite{Voloshin01,Adams03c} in terms of dynamical 
net charge fluctuations measure, 
$\nu_{+-{\rm,dyn}}$, which is found to be a robust observable i.e., independent 
of detection efficiency. $\nu_{+-{\rm,dyn}}$ is defined by the
 expression:

\begin{equation}
\nu _{ +  - ,dyn}  = \frac{{\left\langle {N_ +  (N_ +   - 1)} \right\rangle 
}}{{\left\langle {N_ +  } \right\rangle ^2 }} + \frac{{\left\langle {N_ -  (N_ -   
- 1)} \right\rangle }}{{\left\langle {N_ -  } \right\rangle ^2 }} - 
2\frac{{\left\langle {N_ -  N_ +  } \right\rangle }}{{\left\langle {N_ -  } 
\right\rangle \left\langle {N_ +  } \right\rangle }}
\end{equation}

where $N_{\pm}$ are the number of positively and negatively charged particles in 
the acceptance of interest. Note that 
there exists a 
simple relationship between $\omega_Q$ and $\nu _{ +  - ,dyn}$ written $\nu _{+-
,dyn}=4(\omega_Q-1)/N_{ch}$. This 
relationship is however applicable only if $\omega_Q$ is corrected for finite 
detection effects. Because such 
corrections are not trivial, we favor the use of $\nu _{ +  - ,dyn}$. We note 
additionally that both 
$\omega_Q$ (corrected for efficiency) and $\nu _{ +  - ,dyn}$ may be expressed 
(at least approximately) in terms 
of the variable $D \sim N_{ch}\langle\!\Delta R^2\!\rangle$ 
(with $R=N_{+}/N_{-}$)  used by Koch {\em et al.} \cite{JeonKoch00} for their 
various predictions. Their use, 
for experimental measurements, avoid pitfalls associated with measurements of 
average values of the ratio, $R$, of 
particle multiplicities $-$ where the denominator, $N_{-}$ may be small or even 
zero \cite{AdcoxPRC89}. The 
measurements performed by the STAR \cite{Adams03c} and PHENIX \cite{AdcoxPRC89} 
collaborations 
showed the dynamical net charge fluctuations in $Au + Au$ at $\sqrt{s_{NN}}=$ 
130 GeV are finite but small relative to
 the predictions by Koch {\em et al.} \cite{JeonKoch00} for a QGP. The magnitude 
of the net charge fluctuations 
was found to be in qualitative agreement with HIJING predictions \cite{HIJ} 
although the data exhibit centrality 
dependence not reproduced within the HIJING calculations.  Measured values also 
qualitatively agree with 
predictions by Bialas for quark coalescence \cite{Bialas02} and Koch {\em et 
al.} for a resonance gas 
\cite{JeonKoch00}. 

The scenario for dramatically reduced fluctuations as an evidence for the 
formation of QGP is clearly excluded 
by the data at 130 GeV. However, in light of predictions of a tri-critical point 
of the equation of state 
in the range $10 \le \sqrt{s_{NN}} \le 60$ GeV 
\cite{RajagopalShuryakStephanov,Stephanov:1998dy}, one might argue 
the reduction of fluctuation might be larger at lower beam energies. 
Conversely, one may also argue the volume of QGP formed in $Au + Au$ collisions 
might increase at higher beam 
energies leading to reduced fluctuations at higher beam energy instead. One is 
thus led to wonder whether the 
fluctuations may be found to vary with beam energy thereby indicating the 
production of QGP above a critical 
threshold, or with progressively increasing probability at higher energies. In 
this paper, we consider this 
possibility by investigating how the strength of the dynamical net charge 
fluctuations vary with beam energy and 
system size in $Au + Au$ and $Cu + Cu$ collisions ranging in center of mass 
energy from the highest SPS energy 
to the highest RHIC energy, and relative to $p + p$ collisions at $\sqrt{s_{NN}} 
=$ 200 GeV. The analysis presented 
also provides, independent of existing models, new information that may shed 
light on the collision dynamics.

Various issues however complicate the measurement and interpretation of net 
charge fluctuations. First, one must 
acknowledge that particle final state systems produced in heavy ion collisions 
although large, are nonetheless 
finite and therefore subject to charge conservation effects. Produced particles 
are also measured in a finite 
detector acceptance. Second, one may question whether the dynamical net charge 
fluctuations produced within the
 QGP phase may survive the hadronization process \cite{RonLindenbaum}.  Shuryak 
and Stephanov \cite{Shuryak01} have argued based on 
solutions of the diffusion equation within the context of a model involving 
Bjorken boost invariance, that diffusion
 in rapidity space considerably increases the net charge fluctuations. They 
further argued that the reduced 
fluctuations predicted for a QGP might be observable only if fluctuations are 
measured over a very large rapidity 
range (of order of 4 units of rapidity). Unfortunately, charge conservation 
effects increase with the rapidity 
range considered and might become dominant for rapidity ranges of four units or 
more. Gavin {\em et al.} 
\cite{AbdelAziz05} however argued that the classical diffusion equation yields 
non-physical solutions in the context 
of relativistic heavy ion collisions. They proposed a causal diffusion equation 
as a substitute of the classical 
diffusion equation for studies of net charge fluctuation dissipation. They found 
that causality substantially 
limits the extent to which diffusion can dissipate these fluctuations. 

Third, there exists the possibility that the treatment by Koch {\em et al.} 
\cite{JeonKoch00} of quark and gluons 
behaving as independent particles carrying full entropy may be inappropriate.  
Consider for instance that recent 
measurements 
of elliptical anisotropy of particle emission in $Au + Au$ collisions show that 
meson and baryon elliptical flow, 
$v_2$, scales in proportion to the number of constituent quarks for transverse 
momenta in the range 1-4 GeV/c, thereby
 suggesting that hadrons are produced relatively early in the collisions through 
``coalescence'' or recombination of 
constituent quarks.  In a constituent quark scenario, the role of gluons in 
particle production is reduced. 
Relatively smaller charged particle multiplicities are therefore expected, and 
net charge fluctuations are 
correspondingly larger. Bialas \cite{Bialas02} conducted a simple estimate of 
such a scenario, and reported net 
charge fluctuations $D$ may be of order 3.3.  Interestingly, this estimate 
suggests fluctuations might be even 
larger than that expected for a resonance gas, and as such should also be 
identifiable experimentally. 

Theoretical estimates of the effect of hadronization on net charge fluctuation 
have been for the most part 
restricted to studies of the role of resonances, diffusion 
\cite{Shuryak98,Shuryak01,AbdelAziz05}, and 
thermalization \cite{Gavin04,Gavin04a}. One must however confront the notion 
that collective motion of produced 
particles is clearly demonstrated in relativistic heavy ion collisions. Voloshin 
pointed out \cite{Voloshin03} that induced 
radial flow of particles produced in parton-parton collisions at finite radii in 
nucleus-nucleus collisions generate
 momentum-position correlations not present in elementary proton-proton 
collisions. Specifically, the effect of 
radial flow is to induce azimuthal correlations and to modify particle 
correlation strengths in the longitudinal 
direction. Voloshin showed that two-particle momentum correlations 
$\langle\Delta p_{T1},\Delta p_{T2}\rangle$ are in fact 
sensitive to radial velocity profile as well as the average flow velocity. While 
one may not intuitively expect 
net charge fluctuations to exhibit a dramatic dependence on radial flow, 
simulations based on a simple 
multinomial particle production model including resonances such as the 
$\rho(770)$ indicate that net charge 
correlations are in fact also sensitive to radial flow through azimuthal net 
charge correlations \cite{Pruneau05}. 
They may as such be used to complement estimates of radial velocity obtained 
from fits of single particle spectra 
with blast-wave parameterization or similar phenomenologies.

Measurements of charged particle fluctuations have also been proposed as a tool 
to discriminate between predictions 
of various microscopic models of nuclear collisions. Zhang {\em et al.} 
\cite{Zhang02} find that measurements of 
dynamical fluctuations should exhibit sensitivity to rescattering effects based 
on 
calculations without rescattering with models VNIb \cite{GeigerMuller} and RQMD 
\cite{SorgeStocker}. They also found 
that models VNIb, HIJING \cite{HIJ}, HIJING/B{B} \cite{VanceGyulassy} 
and RQMD predict qualitatively different dependencies on collision centrality. 
Similar conclusions are obtained by
 Abdel-Aziz \cite{AbdelAziz}. 

Bopp and Ranft \cite{Bopp01} compared predictions of net charge fluctuations (at 
mid rapidities) by the dual 
parton model and statistical (thermal) models, and found significant differences 
in the dispersion of the charges 
predicted by these models. They hence argued that charged particle fluctuations 
should provide a clear signal of the 
dynamics of heavy ion processes, and enable a direct measurement of the degree 
of thermalization reached in heavy 
ion collisions.  Gavin \cite{Gavin04,Gavin04a} similarly argued, based on data 
by PHENIX \cite{AdcoxPRC89,AdcoxPR66}
 and STAR \cite{Westfall03,Adams03c} that measured transverse momentum and net 
charge fluctuations indeed 
present evidence for thermalization at RHIC. 

In this work, we present measurements of dynamical net charge fluctuations in 
$Au + Au$ collisions at 
$\sqrt{s_{NN}} =$ 19.6, 62.4, 130, and 200 GeV, $Cu + Cu$ collisions at 
$\sqrt{s_{NN}} =$ 62.4, 200 GeV and in 
$p + p$ collisions at $\sqrt{s} =$ 200 GeV. We study the beam energy, system 
size and collision centrality 
dependencies quantitatively in order to identify possible signature of the 
formation of a QGP. Some of the 
results presented in this work have been reported as preliminary data at various 
conferences 
\cite{Pruneau05}.
The paper is organized into sections on Experimental Method, Results, Systematic 
Error Studies, and Conclusions. 

\section{Experimental Method}

Our study of dynamical net charge fluctuations dependence on the beam energy is 
based on the observable 
$\nu_{+-{\rm,dyn}}$ used in the first STAR measurement \cite{Adams03c}. The 
definition of $\nu_{+-{\rm,dyn}}$, its properties, 
and relationships to 
other measures of event-by-event net charge fluctuations were motivated and 
presented in detail in Refs. 
\cite{Voloshin01,Pruneau02}. The robustness of $\nu_{+-{\rm,dyn}}$ as an 
experimental observable was also discussed 
on the basis of Monte Carlo toy models by Nystrand {\em et al.} 
\cite{Nystrand03}; the authors verified 
explicitly with simple Monte Carlo generators that $\nu_{+-{\rm,dyn}}$ is 
insensitive to the details of the 
detector response and efficiency. Indeed, they verified that values of $\nu_{+-
{\rm,dyn}}$ are independent of the 
track detection efficiency when the efficiency is uniform over the measured 
kinematic range. If the efficiency is not 
perfectly uniform across the acceptance, the robustness of $\nu_{+-{\rm,dyn}}$ 
is reduced in principle. However, in this work, 
the acceptance of the measurement is limited to a kinematic range where the 
efficiency is essentially uniform, and such 
effects are, therefore, negligible.  

We here briefly review the definition and essential properties of this 
observable.
Rather than measuring the event-by-event fluctuations of the ratio of positive 
and negative particle multiplicities 
(in a given acceptance), one considers the second moment of the difference 
between the relative multiplicities 
$N_+/\langle N_{+} \rangle$ and $N_{-}/\langle N_{-}\rangle$ as follows

\begin{equation}
\nu _{ +  - }  = \left\langle {\left( {\frac{{N_ +  }}{{\left\langle {N_ +  } 
\right\rangle }} - \frac{{N_ -  }}{{\left\langle {N_ -  } \right\rangle }}} 
\right)^2 } \right\rangle 
\end{equation}

The Poisson limit, $\nu_{+-,stat}$ of this quantity is equal to:

\begin{equation}
\nu _{ +  - ,stat}= \frac{1}{{\left\langle {N_ +  } \right\rangle }} + 
\frac{1}{{\left\langle {N_ -  } \right\rangle }}
\end{equation}

The ``non-statistical'' or ``dynamical'' fluctuations can thus be expressed as 
the difference between the above 
two quantities:

\begin{eqnarray}
\nu_{+-{\rm,dyn}} &=& \nu_{+-} - \nu_{+-,stat} \\ 
             &=& \frac{\left\langle N_+(N_+-1) \right\rangle}{\left\langle N_+ 
\right\rangle^2} + \frac{\left\langle N_-(N_--1)\right\rangle }{\left\langle N_-
\right\rangle^2} \nonumber \\
             & & -2\frac{\left\langle N_+N_- \right\rangle}{\left\langle N_-
\right\rangle \left\langle N_+ \right\rangle}   \nonumber
\end{eqnarray}

From a theoretical standpoint, $\nu_{+-{\rm,dyn}}$ can be expressed in terms of 
two-particle integral correlation 
functions as $\nu _{ +  - ,dyn}  = R_{ +  + }  + R_{ -  - }  - 2R_{ +  - }$, 
where the terms $R_{\alpha\beta}$ are ratios of 
integrals of two and single particle pseudorapidity density functions defined as 
follows :

\begin{equation}
R_{\alpha \beta}  = \frac{{\int {d\eta _\alpha  d\eta _\beta  \frac{{dN}}{{d\eta 
_\alpha  d\eta _\beta  }}} }}{{\int {d\eta _\alpha  \frac{{dN}}{{d\eta _\alpha  
}}\int {d\eta _\beta  \frac{{dN}}{{d\eta _\beta  }}} } }} - 1
\end{equation}

The dynamical net charge fluctuations variable $\nu_{+-{\rm,dyn}}$ is thus 
basically a measure of the relative 
correlation strength of $++$, $--$, and $+-$ particles pairs. Note that by 
construction, these correlations are 
identically zero for Poissonian, or independent particle production. In 
practice, however, produced particles are 
partly correlated, either through the production of resonances, string 
fragmentation, jet fragmentation, or other 
mechanisms. The relative and absolute strengths of $R_{++}$, $R_{--}$, and 
$R_{+-}$  may vary with colliding 
systems, and beam energy. In addition, by virtue of charge conservation, the 
production of $+-$ pairs is expected to
be more strongly correlated than the production of $++$ or $--$ pairs. For this 
reason, it is reasonable to expect $R_{+-}$ to be larger than $R_{++}$ or $R_{--
}$. In fact, one finds 
experimentally that $2R_{+-}$ is actually larger than the sum $R_{++} + R_{--}$ 
in $p + p$ and $p + \overline{p}$ 
collisions measured at the ISR and FNAL \cite{Whitmore76,Foa75}. Measurements of 
$\nu_{+-{\rm,dyn}}$ are thus expected 
and have indeed been found to yield negative values in nucleus-nucleus 
collisions as well \cite{Adams03c}.

We also note $\nu_{+-{\rm,dyn}}$ is essentially a measure of the variance of 
$N_+ - N_-$. 
This difference is ``orthogonal" to the multiplicity $N_+ + N_-$, and thus 
linearly independent. There is, therefore, 
no bias introduced by binning $\nu_{+-{\rm,dyn}}$ measurements on the basis of 
the reference multiplicity (multiplicity 
within $|\eta| < $0.5) as discussed below.

As a technical consideration, our study of the $\nu_{+-{\rm,dyn}}$ dependence on 
collision centrality is carried out 
in terms of charged particle multiplicity bins, as discussed in detail below. To 
avoid 
dependencies on the width of the bins, we first determine the values of 
dynamical fluctuation,
 $\nu_{+-{\rm,dyn}}(m)$, for each value of multiplicity, {\em m}.  The dynamical 
fluctuations are then averaged across the 
selected finite width of the centrality bins with weights corresponding to the 
relative cross section, $p(m)$, 
measured at each value of multiplicity. For example, in the multiplicity range 
from $m_{min}$ to $m_{max}$, we 
calculate the average as follows:

\begin{equation}
\nu_{+-{\rm,dyn}}(m_{min}\le m < m_{max}) = \frac{\sum{\nu_{+-{\rm,dyn}}(m) 
p(m)}}{\sum{ p(m)}}
\end{equation}

This study is based on the notion that if
 $Au + Au$ collisions (or any other $A + A$ system) trivially consist of a 
superposition of independent 
nucleon-nucleon collisions, with no rescattering of the produced secondaries, 
then $\nu_{+-{\rm,dyn}}$ is expected to 
scale inversely to the number of participating nucleons and the number of 
charged particles, or more appropriately, 
the number of actual nucleon + nucleon collisions. One can thus infer that the 
quantity $|\nu_{+-{\rm,dyn}}dN_{ch}/d\eta|$ 
should be independent of collision centrality under such a scenario. We shall 
therefore examine whether indeed the 
dynamical net charge fluctuations scale with the number of participants, or the 
invariant multiplicity. 

The data used in this analysis were measured using the Solenoidal Tracker at 
RHIC (STAR) detector during the 
2001, 2002, 2004 and 2005 data RHIC runs at Brookhaven National Laboratory. They 
include $Au + Au$ collisions data 
collected at $\sqrt{s_{NN}} =$ 19.6, 62.4, 130, and 200 GeV, $Cu + Cu$ 
collisions data at 
$\sqrt{s_{NN}} =$ 62.4, 200~GeV and $p + p$ collisions data measured at 
$\sqrt{s} =$ 200~GeV. 
The $Au + Au$ collisions at $\sqrt{s_{NN}} =$ 62.4, 130, and 200 GeV data and 
$Cu + Cu$ collisions at 
$\sqrt{s_{NN}} =$ 62.4 and 200 GeV were acquired with minimum 
bias triggers accomplished by requiring a coincidence of two Zero Degree 
Calorimeters (ZDCs) located at 18 m from 
the center of the interaction region on either side of the STAR detector. For 
19.6 GeV data, a combination of minimum 
bias and central triggers was used. The centrality trigger was achieved using a 
set of scintillation detectors, 
called the Central Trigger Barrel (CTB) surrounding the main Time Projection 
Chamber (TPC). Technical descriptions of 
the STAR detector and its components are published in technical reports 
\cite{Star03,Anderson03}. 
For $p+p$ collisions a minimum bias trigger was used based on the CTB detector.
 The analysis carried out in this work is 
rather similar to that published in the first net charge fluctuation measurement 
\cite{Adams03c}.

This analysis is based on charged particle track reconstruction measurements 
performed with the STAR-TPC. 
The TPC is located in a large solenoidal magnetic field producing a uniform 
axial magnetic
 field. The magnetic field was set to 0.25~T for $Au + Au$ collisions at 
$\sqrt{s_{NN}} =$ 19.6 and 130 GeV data, 
and 0.5 T for $Au + Au$ and $Cu + Cu$ collisions at 62.4, and 200 GeV data. The 
increased magnetic field results
 in a slight reduction of the detection efficiency for charged particle tracks 
with transverse momenta below 0.2 
GeV/c, and a modest improvement in momentum resolution. This analysis used 
tracks from the TPC with transverse 
momentum in the range $0.2  < p_T < 5.0$ GeV/c with pseudorapidity $|\eta| 
< 0.5$. Systematic effects associated 
with finite thresholds and momentum dependent efficiency are discussed in 
Section IV. 

In order to limit the net charge fluctuations analysis to primary charged 
particle tracks only (i.e. 
particles produced by the collision), tracks were selected on the basis of their 
distance of closest approach (DCA) 
to the collision vertex. DCA is defined as the distance between the track and 
the primary 
vertex position. A nominal cut of DCA $< 3$ cm was used for results presented 
in this paper.  Systematic effects 
associated with this cut are discussed in Section IV.  

Events were selected for analysis if their collision vertex lay within a maximum 
distance from the center of the 
TPC and they passed a minimum track multiplicity cut (see below). The vertex 
position was determined using a fit involving all 
found tracks. The maximum distance along the beam axis from the center of the 
TPC (also called the z vertex cut) was set
 to 75 cm for the $Au + Au$  19.6 and 130 GeV data, 
further restricted to 25~cm for 62.4 and 200 GeV data. However, a z vertex cut 
of 30 cm was used in $Cu + Cu$ 62.4 
and 200 GeV data. A maximum of 75 cm was used for 
the $p + p$ data. The wide 75 cm cut was used to maximize the event sample used 
in this analysis. The observable 
$\nu_{+-{\rm,dyn}}$  measured in this analysis (as described below) is a robust 
experimental variable, and is by 
construction largely insensitive to restricted detection efficiencies provided  
those efficiencies do not vary 
dramatically across the detector acceptance. We indeed find that as long as the 
longitudinal cut is limited to 
values below 75 cm, for which the track detection efficiency is rather 
insensitive to the pseudorapidity of the 
track, the 
measured values of $\nu_{+-{\rm,dyn}}$ are invariant within the statistical 
uncertainties of the $p+p$ measurements. 
With a larger cut, the efficiencies drop dramatically at large rapidities, and 
$\nu_{+-{\rm,dyn}}$ exhibits somewhat larger 
deviations. The analyses reported in this paper are based on 100k, 1M, 144k, 10M 
$Au + Au$ events at 
19.6, 62, 130, and 200 GeV, respectively, 9M and 5.5M $Cu + Cu$ events at 62 and 
200 GeV, and 2.7M $p+p$ events.

The magnitude of net charge fluctuations is quite obviously subject to change 
with the total multiplicity of 
produced charged particles. It is thus necessary to measure the magnitude of the 
fluctuations and correlations as 
a function of the collision centrality.  Measurements at the AGS, SPS, and RHIC 
have shown that there is a strong anti 
correlation between the number of collision spectators (i.e. projectile/target 
nucleons undergoing little or no 
interaction with target/projectile nucleons) and the multiplicity of charged 
particles produced in the collisions. 
We use the standard collision centrality definition used in other STAR analyses 
and base estimates of
 the collision centrality on the uncorrected multiplicity of charged particle 
tracks measured within the TPC in the 
pseudorapidity range -0.5 $< \eta <$ 0.5. While low multiplicity events 
correspond to peripheral (large impact 
parameter) collisions, high multiplicities are associated with central (small 
impact parameter) collisions. 
The pseudorapidity range -0.5 $<$ $\eta$ $<$ 0.5 is used for collision 
centrality estimates rather than the full 
range -1.0~$<$~$\eta$~$<$~1.0 in principle measurable with the TPC, to 
minimize effects of detector acceptance and 
efficiency on the collision centrality determination. With the narrow cut -
0.5~$<$ $\eta$ $<$~0.5, the track 
detection 
efficiency is rather insensitive to the position of the collision vertex (along 
the beam direction) in the range 
$-75<z<75$ cm used in the analysis of $Au + Au$ at $\sqrt{s_{NN}} =$ 
130 GeV data, and centrality 
selection biases are thus negligible. The efficiency for tracks with 0.5 $<$ 
$\eta$ $< 1$ on the other hand drops 
markedly for vertex positions $|z|>50$ cm. Although the analysis of $Au + Au$ 
and $Cu +Cu$ collisions at 
$\sqrt{s_{NN}} =$ 62.4 and 200 GeV data were conducted with the narrower $|z| 
< 25$ cm and $|z| < 30$ cm range, 
respectively, enabled by the more compact interaction region delivered by the 
accelerator during these runs, the centrality 
determination was estimated on the basis of the same pseudorapidity range in 
order to provide uniform and 
consistent centrality cuts.

The centrality bins were calculated as a fraction of this multiplicity
 distribution starting at the highest multiplicities. The ranges used were 0-5\% 
(most central collisions), 5-10\%, 
10-20\%, 20-30\%, 30-40\%, 40-50\%, 50-60\%, 60-70\%, and 70-80\% (most 
peripheral) for $Au + Au$ collisions. 
Similarly, collision centrality slices used in $Cu + Cu$ collisions are 0-10\% 
(most central), 10-20\%, 20-30\%, 
30-40\%, 40-50\% and 50-60\% (most peripheral). Each centrality bin is 
associated with an average number of 
participating nucleons, $N_{part}$, using Glauber Monte Carlo calculation 
\cite{StarGlauber03}. At low 
multiplicities, the finite detector acceptance and track detection efficiencies 
imply estimates of the collision 
centrality are subject to large errors.  

Events are included or ``counted'' in this analysis provided a collision vertex 
is found (as per the discussion of 
the previous paragraphs) and at least one particle is found in the range $-0.5 
< \eta < 0.5$.  While event counting
 efficiencies are essentially unity for large multiplicity collisions, they are 
limited ($< 1$) for small 
multiplicities corresponding to most peripheral collisions. The limited 
efficiency stems from finite track and 
vertex finding efficiencies. Track finding efficiency within the TPC was studied 
through detailed Monte Carlo 
simulations of the detector response with track embedding. For minimal track 
quality cuts such as those used in 
this analysis, one finds the track finding efficiency is of order 95\% for 
$p_{T}>0.2$ GeV/c in peripheral 
collisions.
 It reduces to approximately 85\% for most central collisions and falls to zero 
for primary tracks with $p_{T} < 0.1$ 
GeV/c. The efficiencies of positive and negative particles are found to be the 
same within the statistical errors. 
The data shown were integrated for tracks with 
$0.2 < p_{T} < 5.0$ GeV/c, $|\eta| < 0.5$ and $0<\phi<2\pi$. Note that 
the minimum $p_{T}$ cut used in this new analysis is different than that used in 
the first reported study 
\cite{Adams03c}. A value of 0.2 GeV/c is used for all measured beam energies and 
field settings to avoid 
systematic effects associated with $p_{T}$ dependent detection efficiency below 
0.2 GeV/c. The results presented 
in this work for 130 GeV are nonetheless in agreement with results reported by 
STAR in the first measurement of net charge 
fluctuations in $Au + Au$ collisions at $\sqrt{s_{NN}} =$ 130 GeV 
\cite{Adams03c}. 

Simulations reveal the vertex finding efficiency is maximum for total charged 
particle multiplicity of 
order 5 and greater in the TPC. We studied the event counting efficiency of this 
analysis with a simple 
simulation based on events generated with the HIJING model \cite{HIJING}, and 
found the event counting efficiency 
is maximum for produced charged particle multiplicities (in the range -0.5 $<$ 
$\eta$ $<$ 0.5) exceeding 12. 
The vertex counting efficiency is of order 90\% for multiplicities larger than 
5, and falls abruptly to zero for smaller 
values. For this reason, the analysis presented in this work is limited to 
reference multiplicities in excess of 10 
and 17 for $Au + Au$ and $Cu + Cu$ collisions where it is deemed minimally 
biased or unbiased.
 
In order to eliminate track splitting we restricted our analysis to charged 
particle tracks producing more than 
20 hits within the TPC where 50\% of these hits were included in the final fit 
of the track.

\section{Net Charge Fluctuation Results}

\begin{figure}[!htb]
\centering
\resizebox{8.7cm}{6.5cm}{\includegraphics{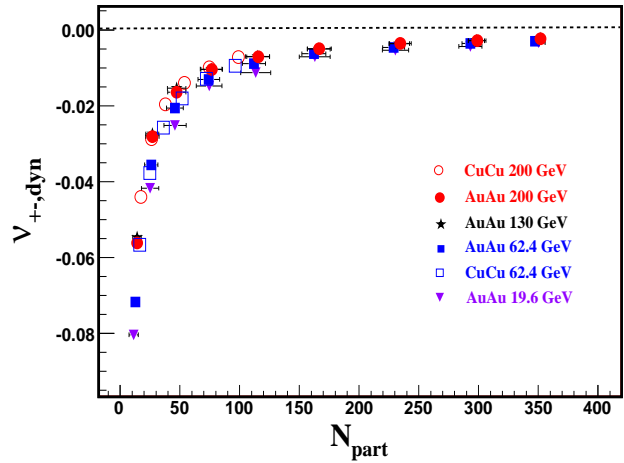}}
\caption[]{(Color online) Dynamical net charge fluctuations, $\nu_{+-
{\rm,dyn}}$, of particles produced within pseudorapidity 
$|\eta|<0.5$, as function of the number of participating nucleons.}
\label{fig1}
\end{figure}

We present, in Fig. \ref{fig1}, measurements of the dynamical net charge 
fluctuations, $\nu_{+-{\rm,dyn}}$, as a 
function of collision centrality in $Au + Au $ collisions at $\sqrt{s_{NN}} =$ 
19.6, 62.4, 130, and 200 GeV, $Cu + Cu$
collisions at $\sqrt{s_{NN}} =$ 62.4 and 200 GeV. 

In Fig. \ref{fig1}, we see that the dynamical net charge fluctuations, in 
general, exhibit a monotonic dependence 
on the number of participating nucleons.
At a given number of participants the measured fluctuations also 
exhibit a modest dependence on beam energy, with $\nu_{+-{\rm,dyn}}$ magnitude 
being the largest in $Au + Au$ collisions 
at $\sqrt{s_{NN}} =$~19.6~GeV. The $\nu_{+-{\rm,dyn}}$ values measured for $p + 
p$ collisions at 
$\sqrt{s} =$ 200 GeV amounts to -0.230 $\pm$ 0.019(stat).

We first discuss the energy dependence of the fluctuations. The collision 
centrality dependence is addressed in 
the following section.

\subsection{Beam Energy and Size Dependence}

A study of the net charge fluctuation dependence on the beam energy is of 
interest given that it can potentially 
reveal a change in the magnitude of the fluctuations and signal the formation of 
QGP. 

We conduct this study primarily on the basis of the 0-5\% and 0-10\% most 
central collisions in $Au + Au$ and 
$Cu + Cu$ collisions, respectively. Extensions to less 
central and peripheral collisions are possible but subject to additional 
uncertainties raised by small systematic 
errors involved the collision centrality determination.
 
As already stated in the Introduction, charge conservation and the finite size 
of the colliding system 
intrinsically limit the magnitude of the net charge correlations. Intuitively, 
one expects charge conservation 
effects to become progressively smaller with increasing charged particle 
multiplicity. Charge 
conservation effects are nonetheless definite at all beam energies and produced 
multiplicities. Specifically, one 
estimates that charge conservation implies a minimum value of order $\nu_{+-
{\rm,dyn}} = -4/N_{4\pi}$, 
where $N_{4\pi}$ is the {\em total} charged particle multiplicity produced over 
$4\pi$ (see \cite{Pruneau02} 
for a derivation of this estimate). This estimate was obtained \cite{Pruneau02} 
assuming that charge conservation 
implies global correlations but no dependence of these correlations on rapidity.  
Therefore, charge conservation effects may be different than those estimated in 
this work.
 Nonetheless, for simplicity, we use the above expression to estimate the 
effects of charge conservation on the 
dynamical net charge fluctuations. 

Corrections to $\nu_{+-{\rm,dyn}}$ for system size and charge conservation 
require knowledge of the total charged 
particle multiplicity. Although, strictly speaking, no experiment at RHIC 
actually measures particle production with 
complete coverage, the PHOBOS experiment comes the closest with a rapidity 
coverage of $|\eta| < 5.4$ over $2 \pi$ 
azimuthal angles and a minimum transverse momentum of order 100 MeV/c. PHOBOS 
has published data on total measured 
charged particle multiplicities of $Au + Au$ collisions at $\sqrt{s_{NN}} = $ 
19.6, 62.4, 130 and 200 GeV 
\cite{Phobos01,Phobos02a,Phobos02b,Phobos04,Phobos05,Phobos06} and $Cu + Cu$ 
collisions at $\sqrt{s_{NN}} =$ 62.4 and 200 
GeV \cite{Phoboscucu}. We infer charged particle multiplicities for $p+p$ 
collisions at 
$\sqrt{s} =$ 200 GeV based on charged particle multiplicity per participant 
reported by PHOBOS \cite{PHOpp}. 
We correct for differences in collision centralities between the PHOBOS and STAR 
measurements using 
a linear interpolation based on the two most central bins measured by PHOBOS. 
Number of participating nucleons 
($N_{part}$), total multiplicities ($N_{ch}$), uncorrected ($\nu_{+-{\rm,dyn}}$) 
and corrected values 
($\nu_{+-{\rm,dyn}}^{corr}$) of $\nu_{+-{\rm,dyn}}$ are shown in Table 
\ref{tab1} 
for $p + p$ collisions at $\sqrt{s} =$ 200 GeV, all four energies in $Au + Au$ 
collisions and two energies 
in $Cu + Cu$ collisions.

\begin{table}[!htp]
\centering{\caption{\label{tab1} Number of participating nucleons, total 
multiplicity, uncorrected and corrected 
$\nu_{+-{\rm,dyn}}$ values for $p + p$ collisions at $\sqrt{s} =$ 200 GeV, four 
energies in $Au + Au$ collisions 
and two energies in $Cu +Cu$ collisions.}}

\begin{tabular}{c||c||c||c||c}
\hline
System \& Energy & $N_{part}$ & $N_{ch}$ & $\nu_{+-{\rm,dyn}}$ & $\nu_{+-
{\rm,dyn}}^{corr}$ \\
\hline \hline
$p + p$ 200 GeV & 2 & 22 & -0.2301 & -0.04407\\
\hline
$Au + Au$ 200 GeV & 351 & 5092 & -0.0024 & -0.00163\\
\hline
$Au +Au$ 130 GeV & 351 & 4196 & -0.0021 & -0.00121\\
\hline
$Au +Au$ 62.4 GeV & 348 & 2788 & -0.0029 & -0.00146\\
\hline
$Au +Au$ 19.6 GeV & 348 & 1683 & -0.0035 & -0.00113\\
\hline
$Cu +Cu$ 200 GeV & 98 & 1410 & -0.0071 & -0.00430 \\
\hline
$Cu + Cu$ 62.4 GeV & 95 & 790 & -0.0093 & -0.00437\\
\hline \hline
\end{tabular}
\end{table}

\begin{figure}[!htp]
\centering
\resizebox{8.7cm}{6.5cm}{\includegraphics{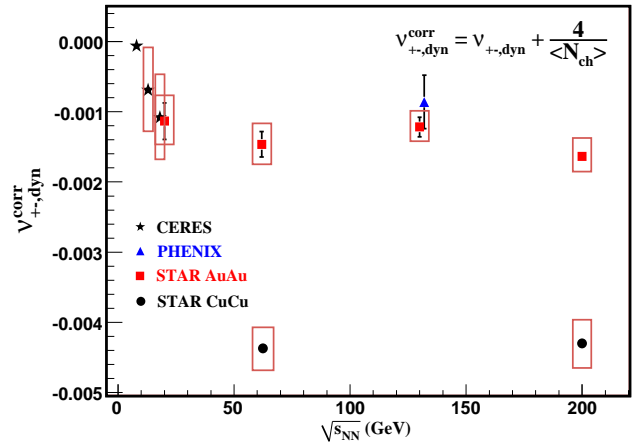}}
\caption[]{(Color online) Corrected values of dynamical net charge fluctuations 
($\nu_{+-{\rm,dyn}}^{corr}$) as a function of 
$\sqrt{s_{NN}}$. See text for details.}
\label{fig2}
\end{figure}

The corrected $\nu^{corr}_{+-,dyn}$ values of the dynamical net charge 
fluctuations 
 are shown in Fig.~\ref{fig2} as function of beam energy for 0-5\% central $Au + 
Au$ collisions with solid squares 
(in red color online) and for 0-10\% central $Cu + Cu$ collisions with solid 
circles (in black online). 
The displayed error bars include (a) the statistical errors involved in the 
measurement of $\nu_{+-{\rm,dyn}}$ 
and (b) the total charged particle multiplicities. The boxes show our estimates 
of the systematic errors involved 
in the measurements of both quantities. Data from this work are compared to 
corrected dynamical net 
charge fluctuations values by the PHENIX and CERES collaborations. The PHENIX 
point (triangle, in blue color online) 
is calculated (as already discussed in \cite{Adams03c}) from data published on 
the basis of the $\omega_Q$ observable 
\cite{AdcoxPRC89} and corrections based on 
total multiplicities measured by PHOBOS (as per values shown in Table 
\ref{tab1}). The CERES data points (star, in 
black online), obtained for $Pb + Au$ collisions, are extracted from their 
published results \cite{CERES04}. 
They include estimates of the systematic errors (open rectangles) as well as 
statistical errors (solid lines).

We first note that the PHENIX and STAR points measured at 130 GeV are in 
quantitative agreement as already 
reported \cite{Adams03c}. The large error bar associated with the PHENIX 
measurement stems mainly from systematic 
uncertainties associated with corrections for detection efficiencies 
\cite{Adams03c}. 
We observe additionally that the STAR 19.6 GeV measurement is in agreement with 
a measurement by CERES at the same 
energy. The STAR measurements in $Cu + Cu$ collisions show a sharp increase in 
magnitude. This difference could partly 
be attributed to the difference in the number of participating nucleons in $Au + 
Au$ and $Cu + Cu$ collisions at 
0-5\% and 0-10\% centralities, respectively. However, the magnitude of 
corrected dynamical fluctuations in $Cu + Cu$ collisions when scaled by the 
ratio of number of participants in 
$Cu + Cu$ collisions to number of participants in $Au + Au$ collisions is -
0.0009 $\pm$ 2$\times 10^{-5}$(stat) $\pm$ 
6$\times 10^{-5}$(sys) and -0.001 $\pm$ 2$\times 10^{-5}$(stat) $\pm$ 8$\times 
10^{-5}$(sys) at $\sqrt{s_{NN}}$ = 62.4 and 200 GeV, 
respectively. We also note that CERES reports 
a dramatic reduction in the magnitude of $\nu_{+-{\rm,dyn}}$ at the lowest 
energy measured at SPS. We thus 
conclude that net charge fluctuations corrected for charge conservation show no 
obvious beam energy dependence 
in the range from 19.6 to 200 GeV. However, there is a clear system size 
dependence when comparing $Au + Au$ to 
$Cu + Cu$ collisions. 

Below 19.6 GeV there appears to be a decrease in the magnitude of $\nu_{+-
{\rm,dyn}}^{corr}$ 
at the lowest SPS energies. Difference between STAR and CERES results may in 
part stem from differences in 
pseudorapidity acceptance.  

Measurements at the SPS have shown that particle production at 5 GeV and lower 
energies is dominated by baryons 
while meson and resonance production become increasingly dominant at energies 
above 19.6 GeV. This suggests that the 
change in dynamical net charge fluctuations below 19.6 GeV might, in part, be 
due to this shift in particle production 
dominance. It is also conceivable that the differences between the values 
measured below and above 19.6 GeV may result from changes 
in the collision dynamics and final state interaction effects 
\cite{RajagopalShuryakStephanov,Stephanov:1998dy,RonLindenbaum,Shuryak01,
AbdelAziz05,Shuryak98,Gavin04,Gavin04a,AbdelAziz,Bopp01}.

\subsection{Collision Centrality Dependence}

The observed monotonic reduction of the magnitude of $\nu_{+-{\rm,dyn}}$ with 
increasing number of participants, seen in 
Fig. \ref{fig1}, arises principally from the progressive dilution of two-
particle correlation  
when the number of particle sources is increased. In fact, one expects $\nu_{+-
{\rm,dyn}}$ to be strictly inversely 
proportional to the number of participating nucleons or the produced particle 
multiplicity if  $Au + Au$ collisions 
actually involve mutually independent nucleon-nucleon interactions, and 
rescattering effects may be neglected. 

\begin{figure}[!htp]
\centering
\resizebox{8.7cm}{6.5cm}{\includegraphics{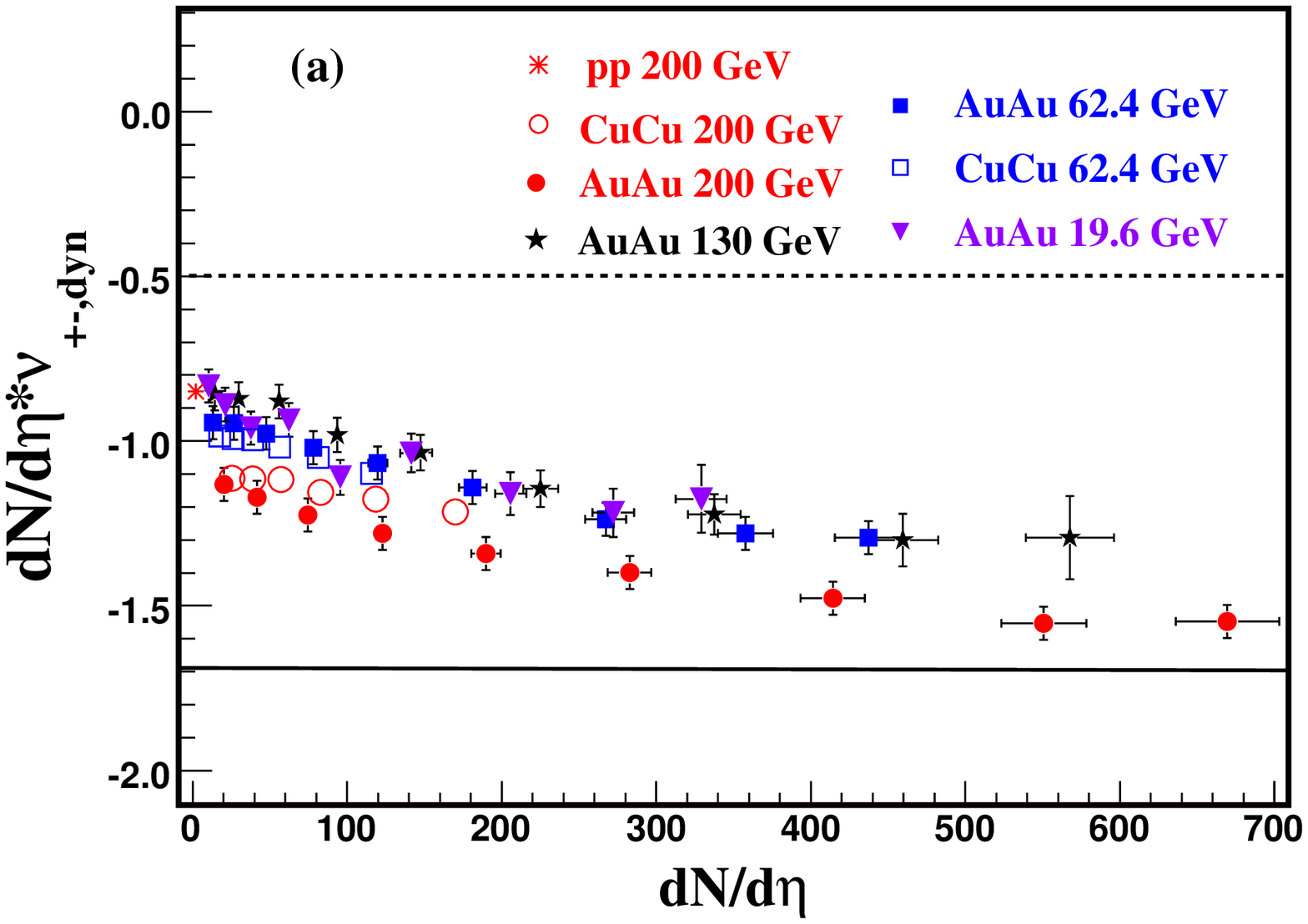}
}
\resizebox{8.7cm}{6.5cm}{\includegraphics{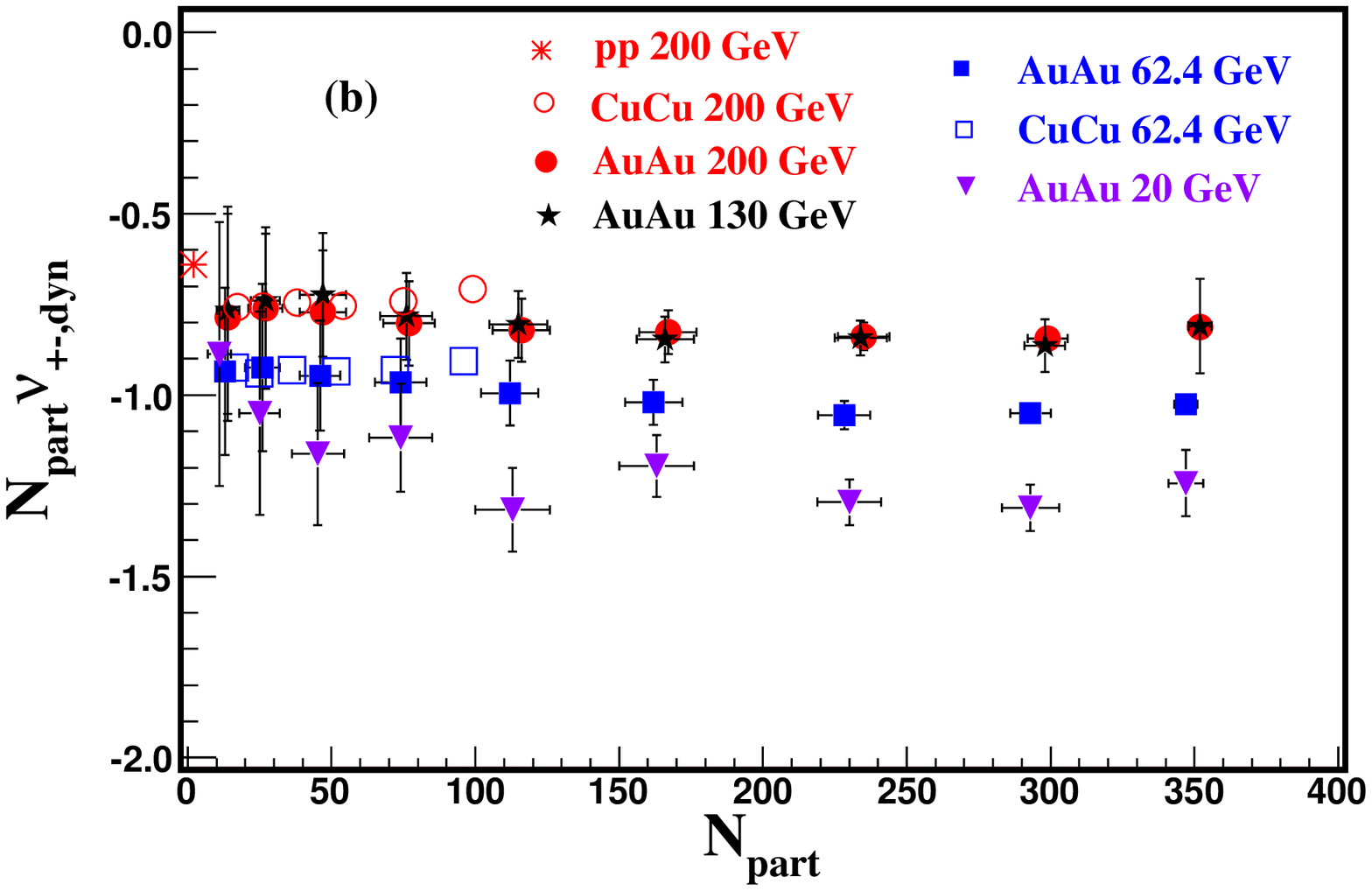}}
\resizebox{8.7cm}{6.5cm}{\includegraphics{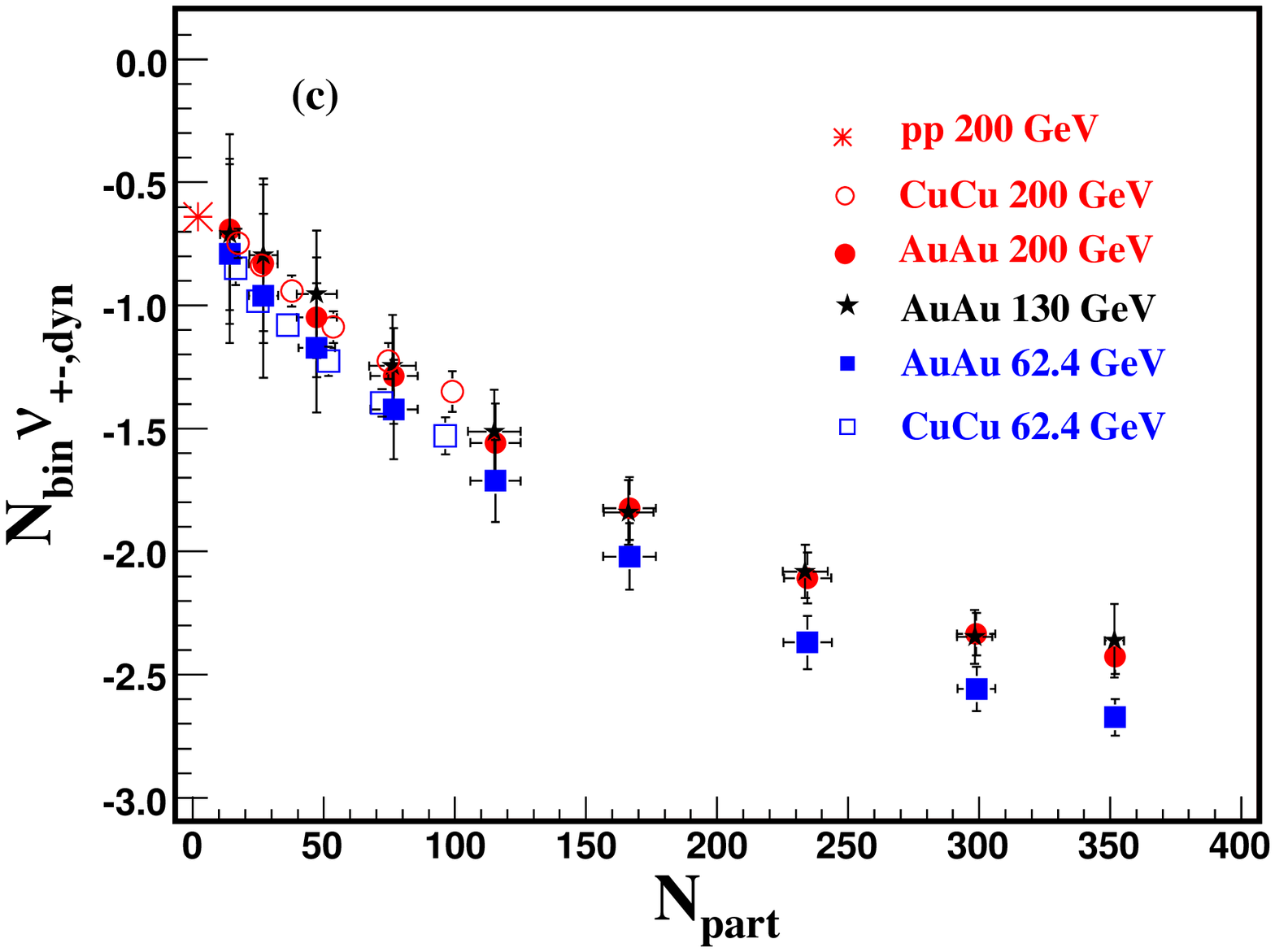}}
\caption[]{(Color online) Dynamical net charge fluctuations, $\nu_{+-
{\rm,dyn}}$, of particles produced with pseudorapidity 
$|\eta|<0.5$ scaled by (a) the multiplicity, $dN_{ch}/d\eta$. The dashed line 
corresponds to charge conservation 
effect and the solid line to the prediction for a resonance gas, (b) the number 
of participants, and (c) 
the number of binary collisions.}
\label{fig3}
\end{figure}

We investigate the possibility of such a scenario by plotting the dynamical 
fluctuations scaled by the measured 
particle multiplicity density in pseudorapidity space ($dN_{ch}/d\eta$) in Fig. 
\ref{fig3}(a). Data from $Au + Au$ 
collisions at various energies are shown with solid symbols while data from $Cu 
+ Cu$ collisions at 62.4 and 200 
GeV are shown with open symbols. Values of $dN_{ch}/d\eta$ used for the scaling 
correspond to efficiency 
corrected charged particle multiplicities measured by STAR \cite{StardNdeta} and 
PHOBOS 
\cite{Phobos01,Phobos02a,Phobos02b,Phobos04,Phobos05,Phobos06,Phoboscucu}. 
We note that the correction applied in Section A to account for charge 
conservation, is useful to study the energy 
dependence of the net charge fluctuations. Its use for centrality, 
pseudorapidity, and azimuthal dependencies is,
however, not warranted given that insufficient data are available to reliably 
account for charge conservation effects. 
Also, the applied correction is model dependent, i.e., assumes charge 
conservation applies only globally \cite{Pruneau02}.

We note from Fig. \ref{fig3}(a) that the magnitude of $\nu_{+-{\rm,dyn}}$ scaled 
by $dN_{ch}/d\eta$ for 
$Au + Au$ 200 GeV 
data is different from the rest of the data. This could partly be attributed to 
the larger multiplicity 
produced in $Au + Au$ 200 GeV. We additionally observe that all four 
distributions exhibit the same qualitative behavior: 
the amplitude $|\nu_{+-{\rm,dyn}}dN_{ch}/d\eta|$ is smallest for peripheral 
collisions, and increases monotonically by $\sim$40\% in central collisions in 
$Au + Au$ and $Cu + Cu$ systems. 
The observed $|\nu_{+-{\rm,dyn}}dN_{ch}/d\eta|$ increases with the increase in 
collision centrality. The
 dashed line in the figure corresponds to charge conservation effect and the 
solid line to the prediction for a 
resonance gas. The figure indicates that dynamical net charge fluctuations, 
scaled by $dN_{ch}/d\eta$ are rather 
large. Most central collisions in $Au +Au$ 200 GeV approach the prediction for a 
resonance gas \cite{JeonKoch00}. 
Indeed, observed values of $\nu_{+-{\rm,dyn}}$ are inconsistent with those 
predicted based on hadronization model of Koch 
{\em et al.} \cite{JeonKoch00,Heiselberg01,Asakawa00}. Given recent observations 
of elliptic flow, suppression of particle 
production at high $p_{T}$ ($R_{AA}$ $\sim$ 0.2), and two-particle 
correlation functions indicating the formation of a strongly interacting medium 
(sQGP) in $A + A$ collisions at RHIC energies, 
this suggests that the signal predicted by the authors 
\cite{JeonKoch00,Heiselberg01,Asakawa00} may be washed out by final state 
interactions, diffusion, expansion, collision dynamics, string fusion \cite{str} 
or other effects 
\cite{RajagopalShuryakStephanov,Stephanov:1998dy,RonLindenbaum,Shuryak01,
AbdelAziz05,Shuryak98,Gavin04,Gavin04a,AbdelAziz,Bopp01,trans100}, some of which were 
discussed in 
the introduction. 
 
Changes in the collision dynamics with increasing centrality are indicated by 
these data. Such a 
conclusion should perhaps not come as a surprise in view of the large elliptical 
flow, and the significant reduction 
of particle production at high transverse momenta reported by all RHIC 
experiments \cite{trans100}. We also note the 
PHOBOS collaboration has reported that the charged particle multiplicity per 
participant nucleon pair rises 
substantially with increasing number of participants. They report a value of 
$dN_{ch}/d\eta/(N_{part}/2)$ of order 
3.9 in central 200 GeV $Au + Au$ collisions compared to a value of 2.5 in $p + 
p$ collisions at the same energy 
\cite{Phobos02b}. This amounts to a 56\% increase, similar in magnitude to that 
of $|\nu_{+-{\rm,dyn}}dN_{ch}/d\eta|$ 
measured in this work. We thus infer that much of the centrality dependence of 
$|\nu_{+-{\rm,dyn}}dN_{ch}/d\eta|$ is due 
to the rise of $dN_{ch}/d\eta/(N_{part}/2)$ with increasing $N_{part}$. 

In order to validate this assertion, we plot in Fig. \ref{fig3}(b) the dynamical 
fluctuation scaled by the 
number of participants, $N_{part} \nu_{+-{\rm,dyn}}$ as a function of the  
number of participants. Vertical error bars 
represent statistical uncertainties. Values of $N_{part} \nu_{+-{\rm,dyn}}$ 
exhibit a small dependence on the collision 
centrality at all four measured energies in $Au + Au$ collisions and two 
energies in $Cu + Cu$ collisions. 
The measured data scaled by the number of participants ($N_{part}$) are thus 
consistent with either no or a very 
weak centrality dependence. 
However, a definite system size and energy dependence is observed. This implies 
that the strength of the 
(integrated) net charge two-particle correlation per participant exhibits 
essentially no dependence on collision centrality.
We also scale $\nu_{+-{\rm,dyn}}$ with the number of binary collisions, shown in 
Fig. \ref{fig3}(c). While we observe that the 
datasets follow a common trend, $\nu_{+-{\rm,dyn}}$ clearly exhibits dramatic 
collision centrality dependence. Such a 
dependence is, however, expected given that the measured dynamical net charge 
fluctuations are dominated by low 
momentum particles with large cross-section for which binary scaling does not 
apply. The statistical errors on 
$\nu_{+-{\rm,dyn}}$ and the scaling factors used in Fig. \ref{fig3}(a), 
\ref{fig3}(b) and \ref{fig3}(c) are added in quadrature.

\subsection{Longitudinal and Azimuthal Dependencies of the Dynamical 
Fluctuations}

Pratt {\em et al.} \cite{Bass00,Jeon01} have argued that the width of 
longitudinal charge balance functions should 
significantly narrow in central $Au + Au$ collision relative to peripheral 
collisions or $p+p$ collisions due to 
delayed hadronization following the formation of a QGP. STAR has in fact 
reported that, as predicted, a narrowing 
of the balance function does occur in central $Au + Au$ collisions relative to 
peripheral collisions 
\cite{StarBalanceFct}. We note, however, 
as already pointed out by Pratt {\em et al.} and more recently by Voloshin 
\cite{Voloshin03}, radial flow 
produced in heavy ion collisions induces large position-momentum correlations 
which manifest themselves in angular, 
transverse momentum, and longitudinal two-particle correlations. The observed 
narrowing of the longitudinal charge 
balance function therefore cannot be solely ascribed to delayed hadronization. 
It is thus important to gauge the 
change in two-particle correlations imparted by radial flow effects. As a first 
step towards this goal, we present 
studies of the net charge fluctuation dependence on the integrated 
pseudorapidity and azimuthal ranges.

\begin{figure}[!htp]
\centering
\resizebox{8.7cm}{6.5cm}{\includegraphics{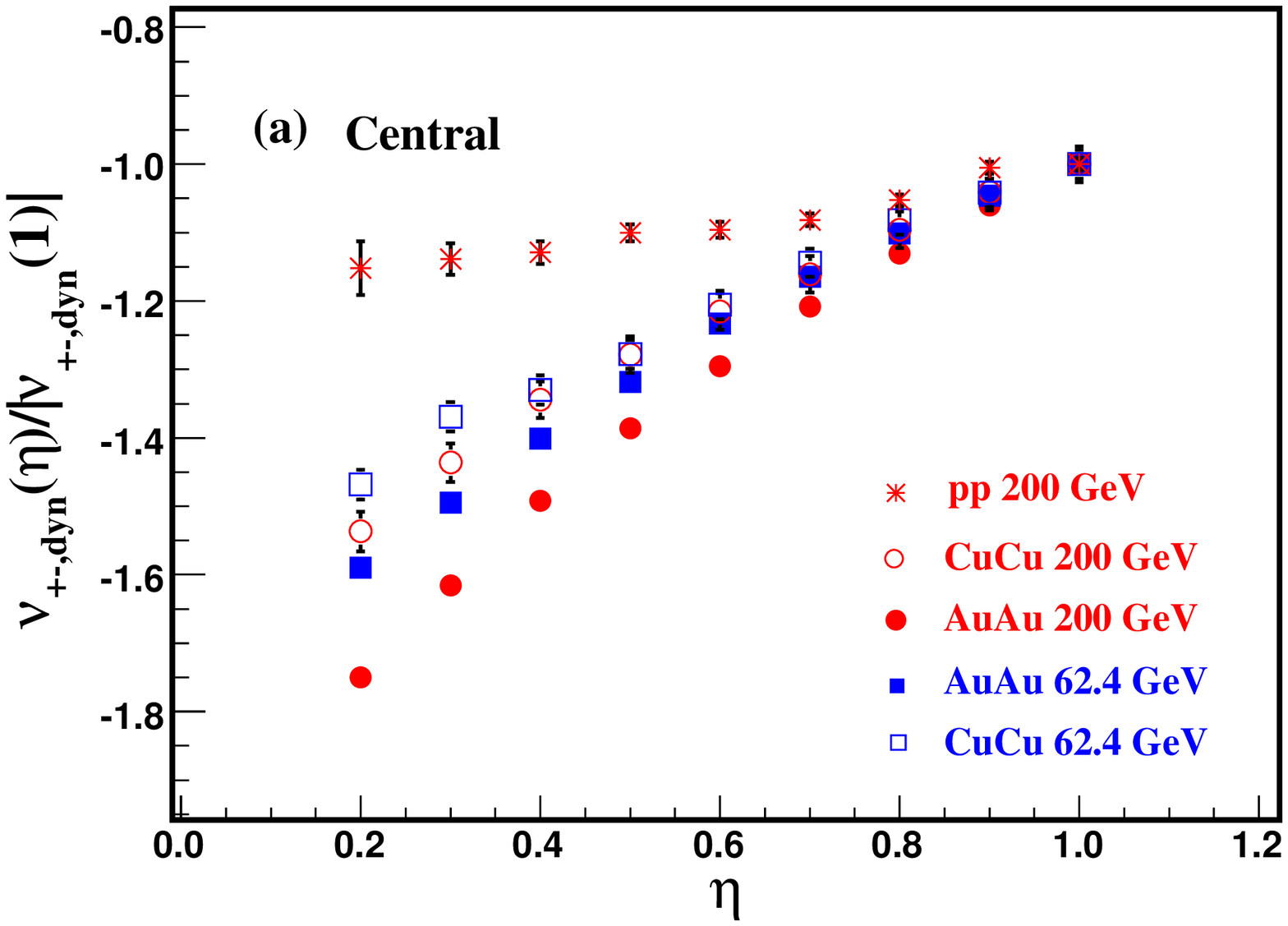}}
\resizebox{8.7cm}{6.5cm}{\includegraphics{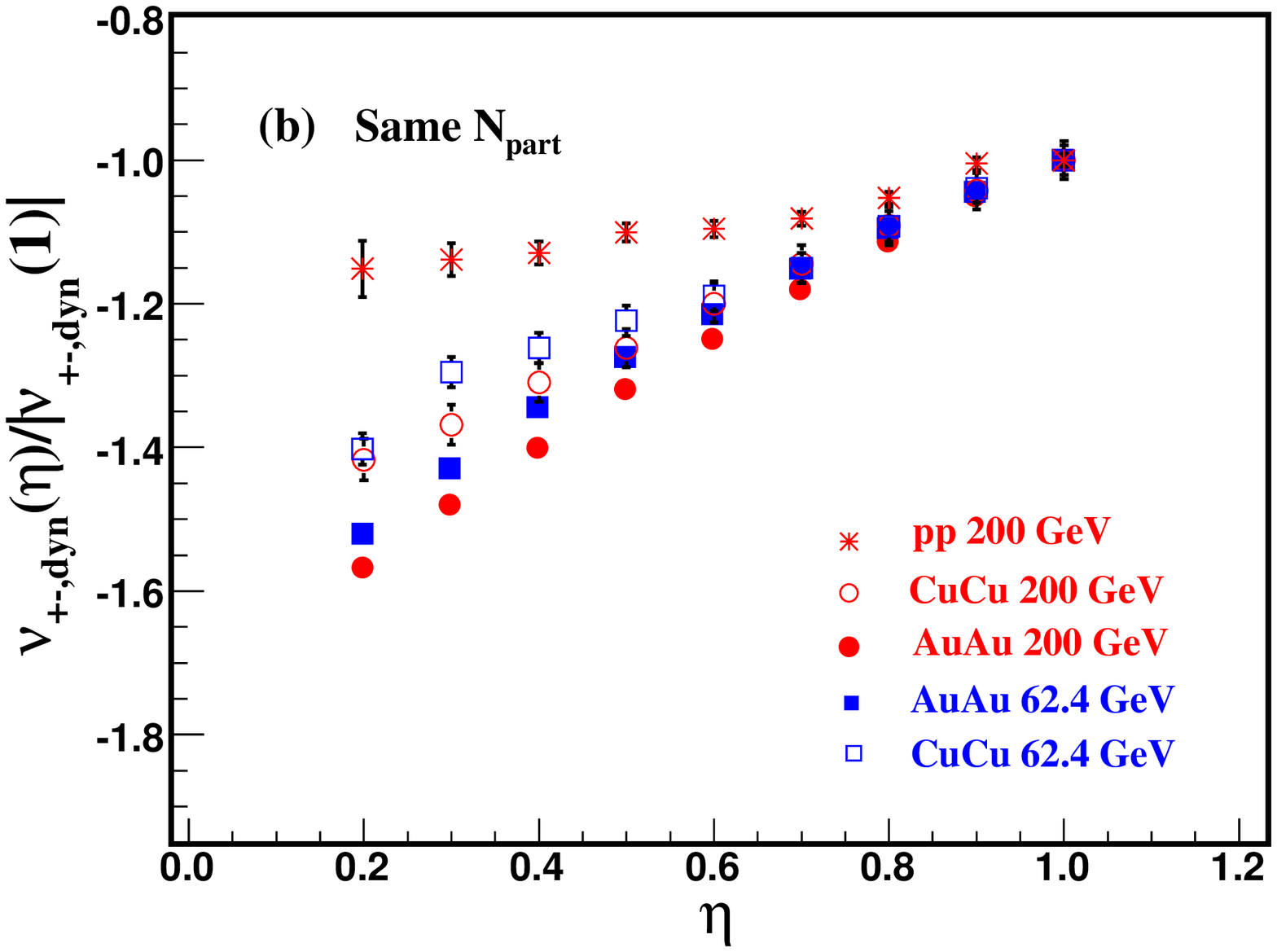}}
\caption[]{(Color online) Dynamical fluctuations $\nu_{+-{\rm,dyn}}$, normalized 
to their value for $|\eta| < 1$, as function of the 
integrated pseudorapidity range. (a) data for $Au + Au$ collisions at 
$\sqrt{s_{NN}}=$ 62.4, 200 GeV 
(0-5\%) along with data for $Cu + Cu$ collisions at $\sqrt{s_{NN}}=$ 62.4, 200 
GeV (0-10\%), are compared to 
inclusive $p + p$ data at $\sqrt{s}=$ 200 GeV, and (b) data for $Au + Au$ 
collisions at $\sqrt{s_{NN}}=$ 
62.4, 200 GeV (30-40\%) along with data for $Cu + Cu$ collisions at 
$\sqrt{s_{NN}}=$ 62.4, 200 GeV 
(0-10\%), are compared to inclusive $p + p$ collision data at $\sqrt{s}=$ 200 
GeV.}
\label{fig5}
\end{figure}

We plot in Fig. \ref{fig5}(a) values of $\nu_{+-{\rm,dyn}}(\eta)$ measured for 
different ranges of pseudorapidity, 
$\eta$. In order to compare data measured at different centralities, beam 
energies and system size, measured values 
are normalized by the magnitude of $\nu_{+-{\rm,dyn}}(\eta)$ for a 
pseudorapidity range $|\eta|<1$ ($\nu_{+-{\rm,dyn}}(1)$). 
The data shown in Fig. \ref{fig5}(a) are from $Au + Au$ collisions at 
$\sqrt{s_{NN}}=$ 62.4 and 200 GeV, $Cu +Cu$ collisions 
at 62.4 and 200 GeV, and $p+p$ data obtained at 200 GeV.
One finds the magnitude of the normalized correlation is maximum for the 
smallest pseudorapidity ranges and 
decreases monotonically to unity, at all energies and centralities, with 
increasing pseudorapidity range.

The dynamical 
fluctuations being essentially a measure of two-particle correlation dominated 
by the $R_{+-}$ term, one finds, 
as expected, that the correlation is strongest for small rapidity intervals, and 
is increasingly diluted (reduced) 
for larger intervals. For example, in $Cu + Cu$ collisions at $\sqrt{s_{NN}}$ = 
200 GeV the typical values of $R_{++}$, 
$R_{--}$ and $R_{+-}$ are 0.99256, 0.992518 and 0.996099, respectively. One 
observes that the magnitudes of 
$|\nu_{+-{\rm,dyn}}(\eta)/\nu_{+-{\rm,dyn}}(1)|$ in $Cu + Cu$ collisions 
at 62.4 and 200 GeV are quite different from $Au + Au$ collisions at comparable 
energies. This shows that the collision 
dynamics  in 
$p + p$ collisions, 0-10\% $Cu + Cu$ and 0-5\% $Au + Au$ collisions are 
significantly different. Indeed, we find 
the relative magnitude of the 
correlations measured for $|\eta|<0.5$ increases by nearly 25\% for $Au + Au$ 
200 GeV relative to those in $p + p$. 
Note in particular that the slope ($d\nu_{+-{\rm,dyn}}/d\eta$) in $p + p$, $Cu + 
Cu$ and $Au + Au$ systems depends on 
the correlation length (in pseudorapidity): the shorter the 
correlation, the larger the slope. The observed distributions then indicate that 
the correlation length is shorter 
for central collisions and for larger systems, in agreement with the observed 
reduction of the charge balance 
function \cite{StarBalanceFct}. The larger values of the slopes observed for 
most central collisions (as well as 
for larger systems) 
indicate correlated pairs of negative/positive particles tend to be emitted 
closer in rapidity than those produced 
in peripheral $Au + Au$ or $p + p$ collisions. Authors of Ref. \cite{Bass00} 
have proposed that a reduction of 
the width of the balance function, and conversely 
a relative increase of short range ($|\eta|<0.5$) correlations, could signal 
delayed hadronization. The 
observed increase in the correlation, reported here, might however also result 
from the strong radial flow believed 
to exist in central $Au + Au$ collisions. 

A comparison of $Au + Au$ collisions at $\sqrt{s_{NN}}=$ 62.4, 200 GeV (30-40\% 
central) is made with 
$Cu + Cu$ collisions at the two energies for 0-10\% centrality in Fig. 
\ref{fig5}(b), as these centralities 
correspond to approximately same number of participant nucleons. We observe that 
the magnitude of normalized correlation 
is similar for both systems at the same beam energy, thereby suggesting that the 
magnitude and the width of the charge 
particle correlation depends mainly on the number of participants and collision 
energy but little on the colliding systems.

To understand the role of radial flow in net charge fluctuations measured in a 
limited azimuthal range (i.e. less
 than $2\pi$), first consider that the magnitude of $\nu_{+-{\rm,dyn}}$ is, in 
large part, determined by the abundance of 
neutral resonances (such as the $\rho(770)$). The decay of neutral resonances 
into pairs of charged 
particles increases the charged particle multiplicity without affecting the 
variance of the net charge. An 
increasing fraction of neutral resonances (relative to other particle production 
mechanisms) therefore leads to 
reduced magnitude of $\nu_{+-{\rm,dyn}}$. Consider additionally that large 
radial flow velocity should lead to a 
kinematical focusing of the decay products in a narrow cone. The opening angle 
of the cone will decrease with 
increasing radial velocity boost. One thus expects that while measuring $\nu_{+-
{\rm,dyn}}$ in a small azimuthal wedge, 
one should have greater sensitivity to the level of kinematical focusing, i.e. 
the magnitude of the 
dynamical net charge fluctuation (correlation) should increase with the 
magnitude of the radial flow velocity. 
Azimuthal net charge correlations should therefore be rather sensitive to the 
magnitude of the radial flow velocity. 

\begin{figure}[!htb]
\centering
\resizebox{8.7cm}{10.5cm}{\includegraphics{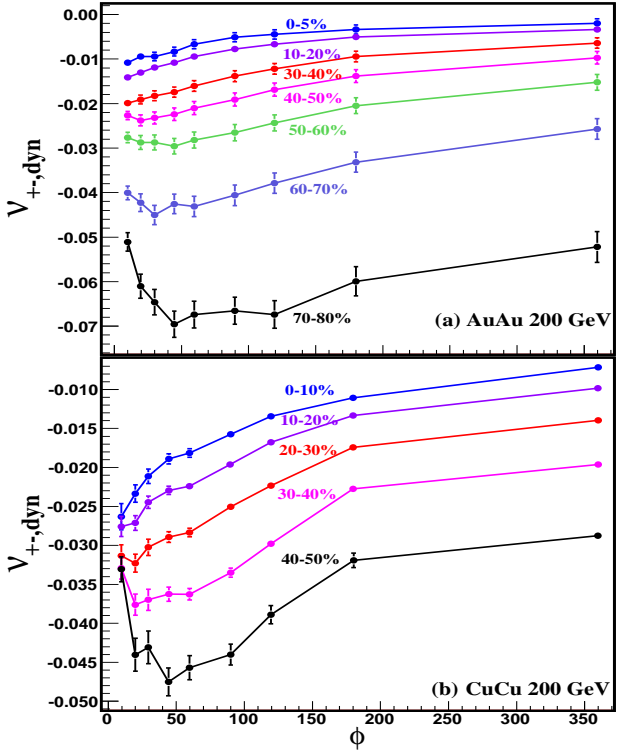}}
\caption[]{(Color online) Dynamical fluctuations $\nu_{+-{\rm,dyn}}$, as a 
function of the integrated azimuthal range $\phi$ for selected
 collision centralities for (a) $Au + Au$ collisions at $\sqrt{s_{NN}}=$ 200 
GeV, and (b) $Cu + Cu$ collisions at $\sqrt{s_{NN}}=$ 200 GeV.}
\label{fig6}
\end{figure}

Fig. \ref{fig6}(a) and (b) display azimuthal net charge correlations integrated 
over azimuthal angle ranges from 
10 to 360 degrees for $Au + Au$ and $Cu +Cu$ collisions at 200 GeV. An azimuthal 
wedge of, for example, 90 degrees would divide 
the complete phase space into four sectors, where we denote each sector as a 
bin. The figure shows results from nine azimuthal 
wedges obtained after averaging $\nu_{+-{\rm,dyn}}$ values for all bins in each 
wedge. The errors shown in \ref{fig6}(a) 
and (b)
 show the statistical errors of the averaged values for each wedge size. We also 
verified that for small wedge angles 
(e.g. 90 degrees and smaller), the variances of the measured values, for wedges 
of a given size, have a 
magnitude similar to the errors of the averages. Data are shown for seven 
collision centrality bins in $Au + Au$ collisions in \ref{fig6}(a) and for 
five centrality bins in $Cu + Cu$ collisions in \ref{fig6}(b). Note that the 
absolute magnitude of the 
correlation decreases from the most peripheral to the central collisions as a 
result of progressive 
dilution with increasing number of participants. The variation of the shape of 
the correlation function with the 
size of the azimuthal acceptance is of greater interest. One finds the 
correlation functions measured in 
the most 
central collisions decrease monotonically in magnitude with increasing azimuthal 
wedge size whereas they exhibit a 
more complicated behavior for most peripheral collisions. One expects $\nu_{+-
{\rm,dyn}}$ to be 
rather small for very small acceptance (azimuthal wedge), i.e., when the size of 
the acceptance is smaller than 
the typical correlation length. This explains why $|\nu_{+-{\rm,dyn}}|$ 
decreases sharply for small angles in peripheral 
collisions. It is remarkable, however, to note that this behavior is not 
observed in most central collisions with 
the angular ranges considered thereby indicating a change in the particle 
correlation length qualitatively 
consistent with the reduction of the balance function in central collision 

already reported by STAR 
\cite{StarBalanceFct}.

\begin{figure}[!htb]
\centering
\resizebox{8.7cm}{8.0cm}{\includegraphics{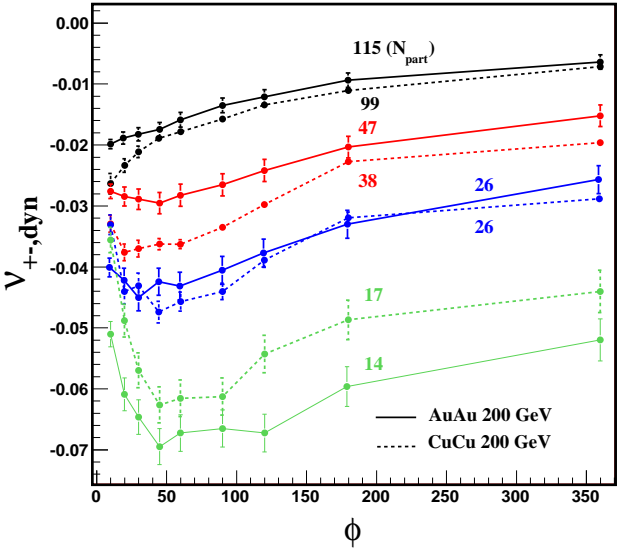}}
\caption[]{(Color online) Dynamical fluctuations $\nu_{+-{\rm,dyn}}$, as a 
function of the integrated azimuthal range $\phi$ for similar
 number of participating nucleons for $Au + Au$ and $Cu + Cu$ collisions at 
$\sqrt{s_{NN}}=$ 200 GeV.}
\label{fig7}
\end{figure}

Fig. \ref{fig7} shows a comparison of $Au + Au$ and $Cu + Cu$ collisions at 
similar number of participating 
nucleons. The magnitude of $\nu_{+-{\rm,dyn}}$ with respect to azimuthal angle, 
$\phi$, is similar for similar number of 
participating nucleons in both systems with best agreement for collisions with 
more 
than 20 participants. The agreement for the most peripheral collisions studied 
is weaker, but we speculate that 
vertex inefficiencies and fluctuations in the number of participants should 
account for this weaker agreement. We 
also observe a change in shape with centrality. Both systems show similar 
monotonic dependence for central collisions, whereas, the magnitude of $\nu_{+-
{\rm,dyn}}$ reaches a maximum for a 
small azimuthal wedge for peripheral collisions. The error bars shown here are 
statistical only. 

\section{ Systematic Error Studies}

While $\nu_{+-{\rm,dyn}}$ is a robust observable and shown to exhibit 
essentially no dependence on efficiencies, 
it may nonetheless be subject to limited systematic effects associated with the 
measurement process. We investigated 
dependencies on the longitudinal position of interaction vertex (z-vertex), the 
effect of resonance feed downs, 
event pile-up, track reconstruction and $p_{T}$ resolution.

The dependence of $\nu_{+-{\rm,dyn}}$ on the longitudinal position of the 
interaction vertex might arise because of the 
restricted acceptance of the TPC on which these analyses are based. We thus 
measured 
$\nu_{+-{\rm,dyn}}$ by binning events according to the z-vertex in steps of 5 cm 
for positions varying in ranges $5<|z|<30$ cm 
and found deviations in $\nu_{+-{\rm,dyn}}$ to be 1\% or less. 

The $\nu_{+-{\rm,dyn}}$ measurement presented in this paper is meant to be 
representative of particles produced by 
$Au + Au$, $Cu + Cu$ or $p + p$ collisions. By design, one thus seeks to 
eliminate effects from secondary decays 
(e.g., $\Lambda \rightarrow p + \pi^{-}$) or secondary particle production 
within the detector. This is accomplished 
by limiting the analysis to tracks that appear to originate from the collision 
vertex. Indeed a cut of track 
distance-of-closest approach (DCA) to the collision vertex with a value of 3 cm 
is used to select primary particles 
and reduce those produced by decays and secondary interactions. The large value 
of DCA used in this analysis is 
due to finite DCA resolution and is also intended to maintain large track 
detection efficiency, which is needed 
especially for the $\nu_{+-{\rm,dyn}}$ 
analysis with respect to the longitudinal and azimuthal acceptance. However, 
with a large value of the DCA cut, one 
ends up counting particles produced by weak-decays (e.g., $\Lambda$ or 
$K_{s}^{0}$) as primary particles. In 
particular, with kaons ($K^{0}_{s}$) representing a small fraction of all 
neutral particles produced, one expects 
pions from decays of these particles to increase the accepted charged particle 
multiplicity but with only a minor impact on the 
variance of the measured net charge. This implies $\nu_{+-{\rm,dyn}}$ should be 
subject to a systematic decrease in 
magnitude when accepting weak-decay feed down. We thus studied $\nu_{+-
{\rm,dyn}}$ for smaller DCA cuts of 2 cm and 
found $|\nu_{+-{\rm,dyn}}|$ decreases by roughly 1\% at all collision 
centralities. 
$K^{0}_{s}$ and $\Lambda$ have a decay length in excess of 2.7 cm. Given the 
rather limited resolution of the 
measurement, the DCA of the decay products is spread to values over a range 
extending more than 3 cm and, thereby, 
form a modest background to the primary particles. Assuming the contributions of 
$K^{0}_{s}$ and $\Lambda$ are roughly 
uniform within the 3 cm DCA cut considered, we expect that a 2 cm DCA cut 
reduces the background by approximately 30\%. 
We observe this change of cut leads to a 1\% reduction in the magnitude of 
$\nu_{+-{\rm,dyn}}$. We thus conclude that 
$K^{0}_{s}$ and $\Lambda$ contamination amounts to a contribution of a few 
percent only.

Another important source of secondary tracks not completely eliminated by the 
DCA cut are electrons/positrons. While a 
finite electron primary yield is expected from decays of D-mesons and B-mesons, 
from Dalitz decays of $\pi^{0}$ and 
$\eta$, the bulk of electrons/positrons observed in the TPC are from secondary 
interactions leading to pair 
production, and Compton photo-electron production. Elimination of 
electrons/positrons is, in principle, partly 
achievable based on cuts on track $dE/dx$. However, because electrons and pions 
of low momenta experience similar 
energy loss in the TPC gas, a cut on the track $dE/dx$ also eliminates a large 
amount of pions thereby effectively 
creating a ``hole'' in the pion acceptance (with respect to their momentum). We 
thus carried out the analysis reported in 
this 
paper by including the electrons/positrons. Again in this case, since electrons 
and positrons are typically created 
in pairs, this may lead to an increase in the integrated charged particle 
multiplicity with little impact on the 
net charge variance. One thus expects inclusion of the electrons should produce 
a systematic shift in the 
magnitude of $\nu_{+-{\rm,dyn}}$. To verify this we carried out a measurement of 
$\nu_{+-{\rm,dyn}}$ when electrons (and
 consequently also pions) are eliminated on the basis of $dE/dx$ cut. The 
$dE/dx$ cut is accomplished using 
the truncated mean of the measured $dE/dx$ samples along the track and the track 
momentum. Tracks 
were excluded whenever the measured $dE/dx$ fell within two standard deviations 
of the mean value expected for 
electrons of a given momentum. We found that when electrons are eliminated, 
$|\nu_{+-{\rm,dyn}}|$ increases by as much 
as 3.5\% in magnitude. This shift may however not be entirely due to the 
suppression of electrons. Indeed, by 
cutting electrons, one also reduces pion acceptance in transverse momentum. We 
have reported in Section III.C. that 
$\nu_{+-{\rm,dyn}}$ exhibit a modest dependence on the size of integrated 
longitudinal and azimuthal acceptances. 
However, a similar (but weaker) dependence on the transverse momentum is 
expected. It is thus plausible the shift 
by 3.5\% may in part result from a reduction of pion acceptance. Electron 
contamination is thus considered a source of 
systematic error of the order of 3.5\% in our measurement of $\nu_{+-
{\rm,dyn}}$. 

$Au + Au$ and $Cu + Cu$ data acquired during runs IV and V were subject to pile-
up effects associated with large
machine luminosity obtained during those years. The pile-up may result in two 
collisions being mistaken as one and 
treated
 as such, thereby leading to artificially large multiplicities and increased 
variances. Therefore, in order to reject 
pile-up events, dip angle cuts were introduced in the present analysis. The dip 
angle is 
defined as the angle between the particle momentum and the drift direction, 
$\theta = \cos^{-1}(p_{z}/p)$. The dip 
angle cut is based on the average dip angle distribution of all tracks in a 
given event. We found the dip angle is 
correlated with the vertex position and features a width distribution which is 
Gaussian at low luminosities. We thus 
reject pile-up events that are beyond two standard deviations of the mean of the 
distribution for a 
particular centrality and vertex position. We found $\nu_{+-{\rm,dyn}}$ changes 
by less than 1\% when the dip angle cut 
is used.  

We also checked the effect of efficiency variation within the acceptance of 
interest. The efficiency is known in 
particular to progressively reduce from a maximum value for $p_{T} > $ 200 MeV/c 
to zero for $p_{T} < $100 MeV/c. 
We determined an upper bound of the effect of $p_{T}$ dependence by measuring 
$\nu_{+-{\rm,dyn}}$ with $p_{T}$ 
thresholds of 150 MeV/c and 200 MeV/c. We found changes of $\nu_{+-{\rm,dyn}}$ 
are typically negligible within the 
statistical accuracy of our measurement and amount to at the most 1.5\%.

Total systematic error contribution increases from 8\% to 9\% from central to 
peripheral 
collisions in $Au + Au$ and $Cu + Cu$ collisions at $\sqrt{s_{NN}} = $ 200 GeV. 
Similarly, systematic errors amount to 
8\% in peripheral collisions and 7\% in central collisions in $Au + Au$ and $Cu 
+ Cu$ collisions at $\sqrt{s_{NN}} = $ 
62.4 GeV.  The systematic errors on $\nu_{+-{\rm,dyn}}$ from different sources 
mentioned above are added linearly.  
\section{Summary and Conclusions}

We have presented measurements of dynamical net charge fluctuations in $Au + Au$ 
collisions at $\sqrt{s_{NN}} =$ 19.6, 
62.4, 130, 200 GeV, $Cu + Cu$ collisions at $\sqrt{s_{NN}} =$ 62.4, 200 GeV and 
$p + p$ collisions at $\sqrt{s} =$ 200 GeV, 
using the measure $\nu_{+-{\rm,dyn}}$. We observed that the dynamical net charge 
fluctuations are non vanishing at 
all energies and exhibit a modest dependence on beam energy in the range 19.6 
$\le \sqrt{s_{NN}} \le$ 200 GeV 
for $Au + Au$ as well as $Cu + Cu$ collisions. Dynamical fluctuations measured 
in this work are in quantitative 
agreement with measurements by the CERES collaboration at $\sqrt{s_{NN}}$ = 17.2 
GeV and PHENIX collaboration at 
$\sqrt{s_{NN}}$ = 130 GeV. However, measurements by CERES at lower beam energy 
($\le$17.2 GeV) exhibit much smaller 
dynamical net charge fluctuations perhaps owing to a transition from baryon to 
meson dominance in the SPS energy 
regime. We also found the dynamical net charge fluctuations violate
 the trivial $1/N_{ch}$ scaling expected for nuclear collisions consisting of 
independent nucleon-nucleon 
interactions. However, one finds that $\nu_{+-{\rm,dyn}}$ scaled by the number 
of participants exhibits little dependence on 
collision centrality but shows modest dependence on collision systems. Measured 
values of $\nu_{+-{\rm,dyn}}$ are inconsistent 
for all systems and energies with the predictions of the QGP hadronization model 
of Koch {\em et al.} \cite{JeonKoch00,Heiselberg01,Asakawa00}. 
Given the reported observations of a strongly interacting medium in $A + A$ 
collisions at RHIC, this suggests that the assumptions of the 
hadronization by Koch {\em et al.} are invalid, or that some final state 
interaction process washes out the predicted signal. 
Scaled dynamical net charge fluctuations $|\nu_{+-{\rm,dyn}}dN_{ch}/d\eta|$ grow 
by up to 40\% from peripheral to 
central collisions. We speculated that the centrality dependence arises, in part 
due to the large radial 
collective flow produced in $Au + Au$ collisions and proceeded to study 
fluctuations as a function of azimuthal angle and pseudorapidity. Our analysis 
showed dynamical fluctuations 
exhibit a strong dependence on rapidity and azimuthal angular ranges which could 
be attributed in part to radial 
flow effects. 

\section*{Acknowledgements}
We thank the RHIC Operations Group and RCF at BNL, and the
NERSC Center at LBNL and the resources provided by the
Open Science Grid consortium for their support. This work 
was supported in part by the Offices of NP and HEP within 
the U.S. DOE Office of Science, the U.S. NSF, the Sloan 
Foundation, the DFG Excellence Cluster EXC153 of Germany, 
CNRS/IN2P3, RA, RPL, and EMN of France, STFC and EPSRC
of the United Kingdom, FAPESP of Brazil, the Russian 
Ministry of Sci. and Tech., the NNSFC, CAS, MoST, and MoE 
of China, IRP and GA of the Czech Republic, FOM of the 
Netherlands, DAE, DST, and CSIR of the Government of India, 
Swiss NSF, the Polish State Committee for Scientific Research,
Slovak Research and Development Agency, and the Korea Sci. 
\& Eng. Foundation.

\vfill\eject
\end{document}